\documentclass{emulateapj}
\usepackage{apjfonts}
\usepackage{comment}
\usepackage{units}
\usepackage{gensymb}
\usepackage{amsmath}
\usepackage{subfigure}
\usepackage{graphicx}

\newcommand{\be}{\begin{eqnarray}}
\newcommand{\ee}{\end{eqnarray}}

\newcommand{\Tsys}{T_{\rm sys}}
\newcommand{\avgdth}{$\langle\Delta\theta\rangle$}

\shorttitle{PALFA Precursor Survey}
\shortauthors{Swiggum et al.}

\begin{document}

\title{Arecibo Pulsar Survey Using ALFA. III. Precursor Survey and Population Synthesis} 

\author{
J.~K.~Swiggum\altaffilmark{1},
D.~R.~Lorimer\altaffilmark{1},
M.~A.~McLaughlin\altaffilmark{1},
S.~D.~Bates\altaffilmark{1,2},
D.~J.~Champion\altaffilmark{3},
S.~M.~Ransom\altaffilmark{4},
P.~Lazarus\altaffilmark{3},
A.~Brazier\altaffilmark{5},
J.~W.~T.~Hessels\altaffilmark{6,7},
D.~J.~Nice\altaffilmark{8},
J.~Ellis\altaffilmark{9},
T.~R.~Senty\altaffilmark{1},
B.~Allen\altaffilmark{9,10,11},
N.~D.~R.~Bhat\altaffilmark{12},
S.~Bogdanov\altaffilmark{13},
F.~Camilo\altaffilmark{13,14},
S.~Chatterjee\altaffilmark{5},
J.~M.~Cordes\altaffilmark{5},
F.~Crawford\altaffilmark{15},
J.~S.~Deneva\altaffilmark{14},
P.~C.~C.~Freire\altaffilmark{3},
F.~A.~Jenet\altaffilmark{16},
C.~Karako--Argaman\altaffilmark{17},
V.~M.~Kaspi\altaffilmark{17},
B.~Knispel\altaffilmark{10,11},
K.~J.~Lee\altaffilmark{18,3},
J.~van~Leeuwen\altaffilmark{6,7}, 
R.~Lynch\altaffilmark{17},
A.~G.~Lyne\altaffilmark{2},
P.~Scholz\altaffilmark{17},
X.~Siemens\altaffilmark{9},
I.~H.~Stairs\altaffilmark{19},
B.~W.~Stappers\altaffilmark{2},
K.~Stovall\altaffilmark{20},
A.~Venkataraman\altaffilmark{14},
W.~W.~Zhu\altaffilmark{19}
}

\altaffiltext{1}{Dept.~of Physics and Astronomy, West Virginia Univ.\, Morgantown, WV 26506, USA}
\altaffiltext{2}{Jodrell Bank Centre for Astrophysics, School of Physics and Astronomy, Univ.\ of Manchester, Manchester, M13 9PL, UK}
\altaffiltext{3}{Max-Planck-Institut f\"ur Radioastronomie, D-53121 Bonn, Germany} 
\altaffiltext{4}{NRAO, Charlottesville, VA 22903, USA} 
\altaffiltext{5}{Astronomy Dept.,  Cornell Univ.\, Ithaca, NY 14853, USA}
\altaffiltext{6}{ASTRON, Netherlands Institute for Radio Astronomy, Postbus 2, 7990 AA, Dwingeloo, The Netherlands} 
\altaffiltext{7}{Astronomical Institute ``Anton Pannekoek'', Univ.\ of Amsterdam, Science Park 904, 1098 XH Amsterdam, The Netherlands}
\altaffiltext{8}{Dept.~of Physics, Lafayette College, Easton, PA 18042, USA}
\altaffiltext{9}{Physics Dept.~ Univ.\ of Wisconsin -- Milwaukee, Milwaukee WI 53211, USA}
\altaffiltext{10}{Leibniz Universit{\"a}t Hannover, D-30167 Hannover, Germany}
\altaffiltext{11}{Max-Planck-Institut f\"ur Gravitationsphysik, D-30167 Hannover, Germany}
\altaffiltext{12}{Center for Astrophysics and Supercomputing, Swinburne Univ.\, Hawthorn, Victoria 3122, Australia} 
\altaffiltext{13}{Columbia Astrophysics Laboratory, Columbia Univ.\,  New York, NY 10027, USA}
\altaffiltext{14}{Arecibo Observatory, HC3 Box 53995, Arecibo, PR 00612, USA}
\altaffiltext{15}{Dept.~of Physics and Astronomy, Franklin and Marshall College, Lancaster, PA 17604-3003, USA} 
\altaffiltext{16}{Center for Gravitational Wave Astronomy, Univ.\ of Texas at Brownsville, TX 78520, USA}
\altaffiltext{17}{Dept.~of Physics, McGill Univ.\, Montreal, QC H3A 2T8, Canada}
\altaffiltext{18}{Kavli institute for radio astronomy, Peking University, Beijing 100871, P. R. China}
\altaffiltext{19}{Dept.~of Physics and Astronomy, Univ.\ of British Columbia, 6224 Agricultural Road Vancouver, BC V6T 1Z1, Canada}
\altaffiltext{20}{Department of Physics and Astronomy, University of New Mexico, NM, 87131, USA}

\begin{abstract}
The Pulsar Arecibo L-band Feed Array (PALFA) Survey uses the ALFA 7-beam receiver to search both inner and outer 
Galactic sectors visible from Arecibo ($32^{\circ}\lesssim \ell \lesssim 77^{\circ}$ and $168^{\circ}\lesssim \ell
 \lesssim 214^{\circ}$) close to the Galactic plane ($|b|\lesssim5^{\circ}$) for pulsars. The PALFA survey is 
sensitive to sources fainter and more distant than have previously been seen because of Arecibo's unrivaled 
sensitivity. In this paper we detail a precursor survey of this region with PALFA, which observed a subset of the 
full region (slightly more restrictive in $\ell$ and $|b|\lesssim1^{\circ}$) and detected 45 pulsars.  Detections 
included one known millisecond pulsar and 11 previously unknown, long-period pulsars. In the surveyed part of the 
sky that overlaps with the Parkes Multibeam Pulsar Survey ($36^{\circ}\lesssim \ell \lesssim 50^{\circ}$), PALFA 
is probing deeper than the Parkes survey, with four discoveries in this region. For both Galactic millisecond and 
normal pulsar populations, we compare the survey's detections with simulations to model these populations and, in 
particular, to estimate the number of observable pulsars in the Galaxy. We place 95\% confidence intervals 
of 82,000  to 143,000 on the number of detectable normal pulsars and 9,000 to 100,000 on the number of 
detectable millisecond pulsars in the Galactic disk. These are consistent 
with previous estimates. Given the most likely population size in each case (107,000 and 15,000 for normal and 
millisecond pulsars, respectively) we extend survey detection simulations to predict that, when complete,
the full PALFA survey should have detected $1,000\substack{+330 \\ -230}$ 
normal pulsars and $30\substack{+200 \\ -20}$ millisecond pulsars. Identical estimation techniques predict that 
$490\substack{+160 \\ -115}$ normal pulsars and $12\substack{+70 \\ -5}$ millisecond pulsars would be detected by 
the beginning of 2014; at the time, the PALFA survey had detected 283 normal pulsars and 31 millisecond pulsars, 
respectively. We attribute the deficiency in normal pulsar detections predominantly to the radio frequency 
interference environment at Arecibo and perhaps also scintillation --- both effects that are currently not 
accounted for in population simulation models.
\end{abstract}


\section{Introduction}\label{sec:intro} 
\setcounter{footnote}{0}
Our current knowledge of the non-recycled (hereafter \emph{normal}) pulsar and millisecond pulsar (MSP) Galactic 
populations\footnote{Although a number of traits separate normal from millisecond pulsars, the most distinct is 
an MSP's short spin period, which is the result of angular momentum transferred by material from a binary companion. 
For the remainder of this paper, we use $P=\unit[30$]{ms} and B$_{\rm surf}=\unit[10^{10}$]{G} as period and surface 
magnetic field thresholds to differentiate between MSPs ($P<\unit[30$]{ms}, B$_{\rm surf}<\unit[10^{10}$]{G}) and 
normal pulsars ($P>\unit[30$]{ms}, B$_{\rm surf}>\unit[10^{10}$]{G}), although there are certainly exceptions to 
this simple separation. A complete list of currently known Galactic MSPs can be found at 
http://astro.phys.wvu.edu/GalacticMSPs} --- their spatial, period and luminosity distributions --- primarily comes 
from the results of the Parkes Multibeam Pulsar Survey \citep[PMPS;][]{man01,mor02,kram03,hob04,fau04,lor06}. 
Analyses of these results have shown that the Galactic normal pulsar population is made up of 30,000$\pm$1,100 
sources beaming toward Earth with luminosities above \unit[0.1]{mJy~kpc$^2$}; their radial density profile is best 
described by a gamma function and their distance from the Galactic plane, by an exponential function with a scale 
height of \unit[0.33]{kpc} \citep{lor06}. A more physically realistic treatment of pulsar luminosities involves 
using a log-normal luminosity function, which is demonstrated from pulsar population syntheses 
\citep[e.g.,][]{fgk06}. The advantage of this approach is that it allows predictions of the \emph{total} normal 
pulsar population size --- not just the number above a certain luminosity cutoff; \cite{fgk06} predict that there 
are 120,000$\pm$20,000 detectable, normal pulsars in the Galaxy.

Since there are only $\sim10\%$ as many known MSPs as normal pulsars \citep{psrcat}, we do not have the same level 
of knowledge about recycled pulsars' population parameters. The High Time Resolution Universe (HTRU) Survey 
\citep{mk10} has added more normal pulsar discoveries to the PMPS haul and many MSPs as well 
\citep[e.g.,][]{bates11,burg13,mor02,hob04,mick12}. Recent analysis of the intermediate latitude portion of HTRU 
MSP detections by \cite{lev13} uses a scale factor method \citep{viv81,lor93} and 50 detected MSPs to place a 
lower limit of 30,000$\pm$7,000 on the Galactic MSP population size (considering sources whose luminosities exceed 
\unit[0.2]{mJy~kpc$^2$}). The scale height of the MSP population is fairly well established to be 500~pc 
\citep{lor05,cor97}, but the spatial, period and luminosity functions are currently less well understood. Although 
many models can be ruled out, plausible MSP populations with a variety of underlying distributions are consistent 
with the observed sample \citep{lor10}.

Despite the fact that Arecibo's latitude does not permit observations close to Galactic center like those at 
Parkes, the unique combination of Arecibo's sensitivity, paired with the high spectral resolution of its back-ends, 
provides a much deeper view through the Galaxy's dispersive medium, which often smears out signals from distant 
sources. Although the PMPS and HTRU surveys have sampled much of the sky surrounding the Galactic center --- an 
area of the sky with high pulsar density --- and have discovered over 1,000 pulsars, PALFA provides a glimpse of 
the population density at larger Galactic radii ($R>\unit[5]{kpc}$), which will help improve the spatial features 
of future pulsar population models. Arecibo's ability to reach competitive sensitivity limits with short 
integration times (1\---5 minutes) makes acceleration searches for binaries unnecessary for all but the most 
exotic systems. Finally, Arecibo's unrivaled sensitivity allows PALFA to probe the low-luminosity end of the 
Galactic pulsar population, leading to a better understanding of the underlying luminosity distribution.

With Arecibo's unique capabilities, PALFA has great potential to discover many normal pulsars as well as MSPs, 
thus improving our statistical picture of each population's characteristics. Given the number of discoveries by 
PMPS, it has historically been used to refine pulsar population modeling assumptions for normal pulsars. Recent 
efforts have been made to discover additional MSPs in archival PMPS data \citep{mick12} with motivation to 
improve MSP population models. With higher sensitivity to dispersed sources and MSPs, the PALFA survey's 
influence on normal and millisecond pulsar population studies will complement those of the PMPS and HTRU surveys. 
MSPs are essential for the direct detection of gravitational waves by pulsar timing array projects 
\citep[e.g.,][]{dem13}. The best way to increase our sensitivity to the stochastic background is to add new MSPs 
to the array \citep{xs13}.

In this paper, we present the detections and discoveries from the initial phase of the PALFA survey, hereafter 
referred to as the ``precursor survey''. In \S\ref{sec:scNda}, we describe the PALFA precursor survey parameters 
and sky coverage and introduce two pipelines used to process the raw data. We present the 45 detections made by 
the precursor survey in \S\ref{sec:surv} and include an evaluation of the survey's efficacy based on measured and 
theoretically calculated signal-to-noise (S/N) ratios. In \S\ref{sec:overlap} we  discuss the portion of sky in the 
precursor survey that overlapped with the PMPS and show preliminary evidence that PALFA will indeed be probing more 
distant, fainter sources. Comparing population simulations to precursor survey detection statistics, we generate 
probability density functions (PDFs) for normal and millisecond pulsar populations in \S\ref{sec:pop}. These PDFs 
inform the predictions we make about the total number of pulsars (normal and MSP) we expect to have detected when 
the full PALFA survey is complete. We conclude in \S\ref{sec:rNd}, stating the most probable normal and millisecond 
pulsar population sizes according to the precursor survey results.

\begin{figure*}[t!]
\begin{center}
\includegraphics[scale=1.0,angle=0]{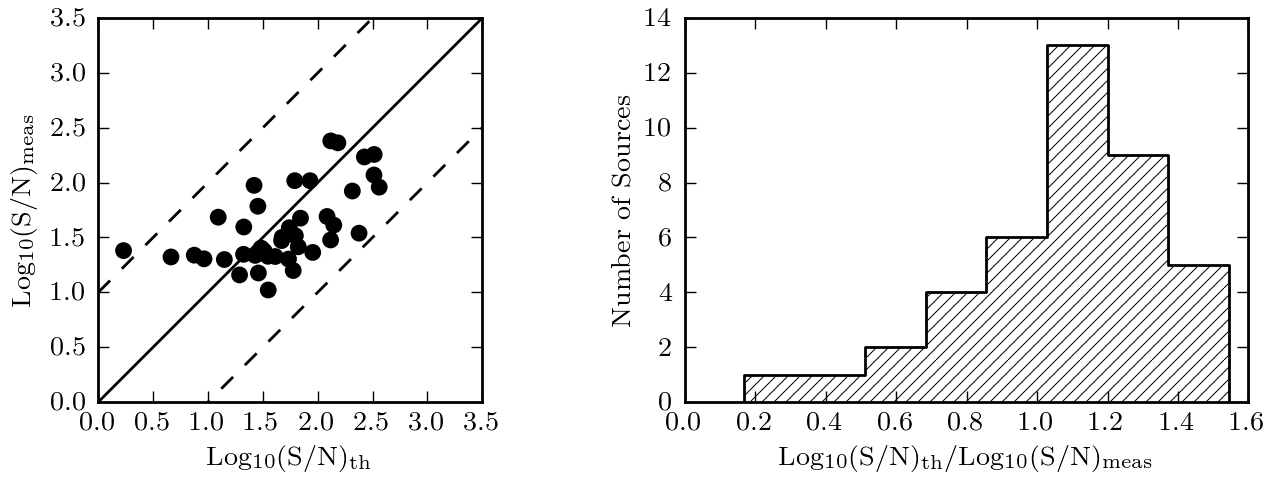}
\caption{The left plot shows theoretical versus measured S/Ns for each source with both quantities available. If the 
two values match for a given source, the data point for that source should lie along the solid line with slope unity. 
The loose correlation shown here is a result of a combination of effects, but most notably, there can be as much as 
$\sim30\%$ fractional error in $({\rm S/N})_{\rm th}$ due to uncertainties in initial flux measurements, which were
taken from the ATNF Pulsar Catalog \citep{psrcat}; interstellar 
scintillation and RFI also contribute to the large scatter. Dashed lines give a reference for sources whose 
theoretical and measured S/N values are different by a factor of 10.
The right plot emphasizes the fact that, in addition to the significant dispersion, $({\rm S/N})_{\rm meas}$ is
smaller than $({\rm S/N})_{\rm th}$ in many cases. This systematic offset implies a poor understanding of the noise
environment and suggests that the maximum sensitivity limits of the survey have not yet been realized.}
\label{fig:snr}
\end{center}
\end{figure*}

\section{Sky Coverage and Data Analysis}\label{sec:scNda}
The PALFA precursor survey covered portions of two Galactic sectors --- an inner Galaxy region, 
$36^{\circ}\lesssim \ell \lesssim 75^{\circ}$, tiled with 865 pointings, and an outer Galaxy region, 
$170^{\circ}\lesssim \ell \lesssim 210^{\circ}$, covered by 919 pointings. All pointings were within one degree of 
the Galactic plane ($|b|<1^{\circ}$) and had dwell times of 134 and 67 seconds for inner- and outer-Galaxy regions 
respectively. The precursor survey used the Arecibo L-band Feed Array (ALFA) 7-beam receiver in conjunction with the 
Wideband Arecibo Pulsar Processor (WAPP) back-end \citep{wapp}, which was set up to record 256 channels covering a 
\unit[100]{MHz} bandwidth, centered at \unit[1.42]{GHz}, every \unit[64]{$\mu$s}. Each ALFA pointing includes seven 
distinct beam positions in a hexagonal pattern. As PALFA continues, the sky coverage will increase slightly in Galactic 
longitude ($32^{\circ}\lesssim \ell \lesssim 77^{\circ}$ and $168^{\circ}\lesssim \ell \lesssim 214^{\circ}$) and will 
extend to Galactic latitude $\pm5^{\circ}$. For the remainder of the paper, we will refer to this extended spatial 
coverage (accompanied by a three-fold increase in bandwidth) as the \emph{full PALFA survey}. The precursor 
survey, optimized for maximum efficiency and sensitivity, used a ``sparse sampling'' technique described in detail 
in \cite{cor06}; gaps left by the precursor survey will be covered in multiple passes by the full PALFA survey.
PMPS overlaps with the southernmost regions covered by Arecibo in the PALFA precursor survey, corresponding to 
$36^{\circ}\lesssim \ell \lesssim 50^{\circ}$. In \S\ref{sec:surv}, we will compare the performance of the two surveys 
in this overlap region to make a statement about the efficacy of the PALFA precursor survey.

Data from the PALFA precursor were previously analyzed in \cite{cor06}.  That analysis used a quasi-real-time 
{\sc Quicklook} pulsar search pipeline in which the data were decimated in time and frequency by factors of 8 and 16, 
respectively, yielding 32 spectral channels and 1024 $\mu$s time resolution.   Using the decimated data, 11 pulsars 
were discovered and 29 previously known pulsars were detected. Timing and spectral characteristics from follow-up 
observations of the newly discovered pulsars are given in \cite{nice13}.

We have analyzed these same data files at native full time- and frequency-resolution using the PALFA survey's 
{\sc Presto 1} pipeline. The full resolution search of the precursor survey data did not yield any pulsar discoveries 
(and in fact missed some sources flagged by the {\sc Quicklook} pipeline), but revealed two more previously 
known normal pulsars (J1946+2611, B1924+16) and the bright MSP~B1937+21. The {\sc Presto 1} zaplist, a list of 
frequencies and their harmonics related to known sources of RFI, may be responsible for this scant improvement
over {\sc Quicklook} results since it was fairly restrictive, ``zapping'' $\sim8\%$ of the 
spectral region between \unit[0$-$10]{Hz} ($\sim84\%$ of known pulsars have spin frequencies in this range).
At least one previously known source, B1925+188, fell inside a zapped portion of the spectrum, but its fourth
harmonic was still detectable in {\sc Presto 1} results. Four other sources that were detected by {\sc Quicklook}
 (J1913+1000, B1919+14, J2002+30 and J2009+3326) were not detectable in {\sc Presto 1} results. Of the 12,488 
PALFA precursor beams, 183 ($1.5\%$) were not processed by the {\sc Presto 1} pipeline, including beams where
J1913+1000 and B1919+14 should have been detected. PSRs J2002+30 and J2009+3326 were processed by
{\sc Presto 1} and their spin frequencies were outside zapped portions of the spectrum; why these two sources were
not detectable remains unknown, although it is plausible that harmonics of their true spin frequencies could have
been ``zapped,'' causing these sources to fall below a detectable threshold.

After the precursor 
survey was complete, raw data products were decimated to 4-bit resolution and saved in that form. In the process, 
some files were lost or corrupted (i.e. detection data files for J1913+1000, B1919+14 and B1924+16), so results 
from \cite{cor06} were used when necessary. We used a complete list of precursor beam
positions to determine minimum offset angles from each known source in the survey region, then refolded corresponding 
4-bit data files, yielding two additional detections (J1906+0649 and J1924+1631). Table \ref{tab:full} outlines the means
by which all sources in the PALFA precursor survey were detected.

\subsection{{\sc Presto 1} Pipeline}\label{sec:prestpipe}
The PALFA {\sc Presto 1} pipeline\footnote{Many of the aspects of the PALFA precursor survey data processing described 
here have since been augmented (e.g., \cite{laz13}), including a new complementary pipeline 
based on the Einstein@Home distributed volunteer computing platform, e.g. \cite{e@h13}.} used to analyze precursor 
survey data first converted WAPP-format data to SIGPROC filterbank-format \citep{lor01}. Each filterbank file, one 
per beam, was then processed independently using various programs from the \textsc{Presto} suite of pulsar analysis 
software\footnote{https://github.com/scottransom/presto} \citep{rans02}. Strong narrow-band impulsive and periodic 
signals were identified as interference by {\tt rfifind}. The filterbank files were then cleaned and 
reduced-frequency-resolution sub-band files were created at various dispersion measures (DMs). Each group of sub-band 
files was then used to create time series with DMs close to the DM of the sub-band file. In total 1056 trial DM values 
were used between $0 \leq {\rm DM} \leq \unit[1003.2$]{pc~cm$^{-3}$}. The upper limit was chosen to reflect the 
maximum expected DM in the sky region surveyed \citep{ne2001}.

Each dedispersed time series was searched for single pulses using {\tt single\_pulse\_search.py}. Significant pulses 
($\sigma > 6$) with widths up to 0.1 s were identified and a diagnostic plot was generated for human inspection. 
The time series were also Fourier transformed and searched for periodic signals using {\tt accelsearch}. The 
periodicity search was done in two parts, one for unaccelerated pulsars using up to 16 summed harmonics and the 
other for accelerated pulsars using up to 8 summed harmonics. The high-acceleration search used a Fourier-domain 
algorithm \citep{rans02} with a maximum drift of 50 FFT bins. Non-pulsar-like signals were removed from the 
candidate lists generated from the low and high-acceleration searches. The manicured low and high-acceleration 
candidate lists were then combined. Candidates harmonically related to a stronger candidate were discarded, while 
the top 50 candidates with $\sigma > 6$ were ``folded'' modulo the best Fourier-detected period using  
 {\tt prepfold}, which effectively provides a fully-coherent harmonic sum of the signal power. 
The resulting plots, along with basic metadata about the observations were loaded into a 
database hosted at Cornell University, where volunteers selected and inspected candidate plots.

\subsection{Detection {\rm S/N} Measurements}\label{sec:snmeas}
For all sources detected by the Quicklook and {\sc Presto 1} processing pipelines, we refolded data files from
beam positions nearest those sources using known pulsar parameters and calculated measured signal-to-noise
$({\rm S/N})_{\rm meas}$ values. For each pulse profile, we used a simple algorithm to determine
on- and off-pulse bins, then summed on-pulse intensities and divided by the maximum profile intensity to get an equivalent
top-hat pulse width $W_{\rm eq}$ (in bins). Finally, $({\rm S/N})_{\rm meas}$ is computed with
\begin{equation}
({\rm S/N})_{\rm meas} =\frac{1}{\sigma_{p}\sqrt{W_{\rm eq}}}\sum_{i=1}^{n_{\rm bins}} (p_i-\bar{p}),
\label{snrmeas}
\end{equation}
as in \cite{lk05}, where $\bar{p}$ and $\sigma_{p}$ are the mean and standard 
deviation of off-pulse intensities respectively, $p_i$ is
the intensity of an individual profile bin and each profile had $n_{\rm bins}=128$. We divided $W_{\rm eq}$ by the number of
bins in a profile $n_{\rm bins}$ to convert to duty cycle $\delta$ for each detection. Computed $\delta$ and 
$({\rm S/N})_{\rm meas}$ values are listed in Table \ref{tab:full}.

\section{Survey Results}\label{sec:surv}
To measure the effectiveness of a pulsar survey, we look at the known sources that fall inside the survey region and 
compare the number of detections to the number of expected detections. Effectiveness will then be evaluated by whether 
the survey meets/exceeds expectations for detecting individual sources.

\begin{deluxetable*}{lccccccccccc}[ht!]
\tablewidth{500pt}
\tablecaption{\label{tab:full}Detections and expected detections by the precursor survey}
\tablecolumns{12}
\tablehead{\colhead{PSR Name}  & \colhead{$P$} & \colhead{DM} &  \colhead{$\ell$} & \colhead{$b$} &
\colhead{$\Delta\theta$} & \colhead{Duty Cycle} & \colhead{Flux Density} & \colhead{$({\rm S/N})_{\rm th}$} & 
\colhead{$({\rm S/N})_{\rm meas}$} & \colhead{Pipeline Detected?} &\colhead{PALFA} \\
& (s) & (pc cm$^{-3}$) & ($\degree$) & ($\degree$) & ($'$) & (\%) & (mJy) & & & (QL / P1 / Refold) & Discovery?}
\startdata
J0540+3207 & 0.524 & 61 & 176.7 & 0.8 & 1.43 & 2.1 & 0.34 & 62.6 & 32.8 & QL, P1, Refold & Yes \\
J0628+0909 & 1.241 & 88 & 202.2 & --0.9 & 2.30 & 1.4 & 0.06 & 4.6 & 21.0 & QL, P1, Refold & Yes \\
J0631+1036 & 0.288 & 125 & 201.2 & 0.5 & 1.51 & 3.3 & 0.80 & 85.1 & 104.1 & QL, P1, Refold & \\
J1855+0307 & 0.845 & 402 & 36.2 & 0.5 & 3.24 & 1.7 & 0.97 & 12.4 & 48.4 & QL, P1, Refold & \\
J1901+0621 & 0.832 & 94 & 39.7 & 0.8 & 1.76 & 5.6 & 0.47 & 35.2 & 21.3 & QL, P1, Refold & Yes \\
B1859+07 & 0.644 & 252 & 40.6 & 1.1 & 2.29 & 3.0 & 0.90 & 55.1 & 38.8 & QL, P1, Refold & \\
J1904+0738 & 0.209 & 278 & 41.2 & 0.7 & 0.90 & 1.9 & 0.23 & 54.2 & 20.1 & QL, P1, Refold & Yes \\
J1904+0800 & 0.263 & 438 & 41.5 & 0.9 & 1.99 & 2.8 & 0.36 & 41.0 & 21.2 & QL, P1, Refold & \\
J1905+0616 & 0.990 & 256 & 40.1 & --0.2 & 1.80 & 1.5 & 0.51 & 69.7 & 47.4 & QL, P1, Refold & \\
B1903+07 & 0.648 & 245 & 40.9 & 0.1 & 0.52 & 5.6 & 1.80 & 266.2 & 171.3 & QL, P1, Refold & \\
& & & & & & & & & & & \\
J1905+0902 & 0.218 & 433 & 42.6 & 1.1 & 0.50 & 1.9 & 0.10 & 21.1 & 22.2 & QL, P1, Refold & Yes \\
B1904+06 & 0.267 & 472 & 40.6 & --0.3 & 2.43 & 5.6 & 1.70 & 61.8 & 104.1 & QL, P1, Refold & \\
J1906+0649 & 1.287 & 249 & 40.7 & --0.2 & 2.53 & 6.3 & 0.30 & 9.2 & 20.2 & Refold &\\
J1906+0746 & 0.144 & 217 & 41.6 & 0.1 & 2.60 & 1.6 & 0.55 & 28.8 & 15.0 & QL, P1, Refold & Yes \\
J1906+0912 & 0.775 & 265 & 42.8 & 0.9 & 2.37 & 2.5 & 0.32 & 19.4 & 14.4 & QL, P1, Refold &\\
J1907+0740 & 0.575 & 332 & 41.6 & --0.1 & 2.24 & 2.2 & 0.41 & 30.5 & 25.3 & QL, P1, Refold & \\
J1907+0918 & 0.226 & 357 & 43.0 & 0.7 & 3.00 & 1.6 & 0.29 & 7.5 & 21.8 & QL, P1, Refold &\\
J1908+0734 & 0.212 & 11 & 41.6 & --0.3 & 1.05 & 3.1 & 0.54 & 90.1 & 23.1 & QL, P1, Refold &\\
J1908+0909 & 0.337 & 467 & 43.0 & 0.5 & 1.70 & 2.2 & 0.22 & 28.5 & 60.9 & QL, P1, Refold &\\
B1907+10 & 0.284 & 149 & 44.8 & 1.0 & 1.92 & 2.3 & 1.90 & 206.9 & 83.8 & QL, P1, Refold &\\
& & & & & & & & & & \\
J1910+0714 & 2.712 & 124 & 41.5 & --0.9 & 1.72 & 1.4 & 0.36 & 59.8 & 15.8 & QL, P1, Refold &\\
B1910+10 & 0.409 & 147 & 44.8 & 0.2 & 2.32 & 3.7 & 0.22 & 11.0 & --- & --- &\\
J1913+1000 & 0.837 & 422 & 44.3 & --0.2 & 1.69 & 3.8 & 0.53 & 66.5 & 26.0 & QL &\\
J1913+1011 & 0.036 & 178 & 44.5 & --0.2 & 2.69 & 4.1 & 0.50 & 14.1 & 19.9 & QL, P1, Refold &\\
J1913+1145 & 0.306 & 637 & 45.9 & 0.5 & 2.06 & 4.7 & 0.43 & 23.4 & --- & --- & \\
B1911+11 & 0.601 & 100 & 45.6 & 0.2 & 1.90 & 4.2 & 0.55 & 43.9 & --- & --- &\\
B1913+10 & 0.405 & 241 & 44.7 & --0.7 & 1.51 & 1.6 & 1.30 & 238.2 & 34.6 & QL, P1, Refold &\\
B1914+13 & 0.282 & 237 & 47.6 & 0.5 & 1.78 & 2.4 & 1.20 & 152.8 & 230.4 & QL, P1, Refold &\\
B1915+13 & 0.195 & 94 & 48.3 & 0.6 & 2.29 & 2.5 & 1.90 & 131.6 & 239.9 & QL, P1, Refold &\\
B1916+14 & 1.181 & 27 & 49.1 & 0.9 & 3.04 & 1.4 & 1.00 & 26.9 & 21.7 & QL, P1, Refold &\\
& & & & & & & & & & &\\
B1919+14 & 0.618 & 91 & 49.1 & 0.0 & 0.45 & 3.6 & 0.68 & 140.2 & 41.0 & QL &\\
B1921+17 & 0.547 & 143 & 51.7 & 1.0 & 3.01 & 3.6 & --- & --- & 46.8 & QL, P1, Refold & \\
J1924+1631 & 2.935 & 518 & 51.4 & 0.3 & 0.65 & 1.0 & 0.09 & 35.4 & 10.5 & Refold & \\
B1924+16 & 0.580 & 176 & 51.9 & 0.1 & 0.83 & 2.5 & 1.30 & 363.5 & 90.9 & P1 &\\
B1925+188 & 0.298 & 99 & 53.8 & 0.9 & 1.92 & 5.9 & --- & --- & 27.8 & QL, P1, Refold &\\
J1928+1746 & 0.069 & 176 & 52.9 & 0.1 & 0.70 & 5.2 & 0.28 & 46.9 & 29.6 & QL, P1, Refold & Yes \\
B1929+20 & 0.268 & 211 & 55.6 & 0.6 & 3.78 & 2.0 & 1.20 & 1.7 & 24.0 & QL, P1, Refold &\\
B1937+21 & 0.00156 & 71 & 57.5 & --0.3 & 2.41 & 14.9 & 13.20 & 327.0 & 180.5 & P1, Refold &\\
J1946+2611 & 0.435 & 165 & 62.3 & 0.6 & 2.61 & 2.4 & --- & --- & 18.5 & P1, Refold &\\
B1952+29 & 0.427 & 7 & 66 & 0.8 & 2.53 & 4.5 & 8.00 & 325.8 & 117.3 & QL, P1, Refold &\\
& & & & & & & & & & &\\
J1957+2831 & 0.308 & 138 & 65.5 & --0.2 & 1.57 & 3.6 & 1.00 & 131.2 & 30.0 & QL, P1, Refold &\\
J2002+30 & 0.422 & 196.0 & 67.9 & --0.2 & 1.21 & 3.7 & --- & --- & 60.7 & QL, Refold &\\
B2000+32 & 0.697 & 142 & 69.3 & 0.9 & 2.16 & 1.8 & 1.20 & 121.7 & 49.1 & QL, P1, Refold &\\
B2002+31 & 2.111 & 234 & 69.0 & 0.0 & 3.30 & 1.3 & 1.80 & 26.3 & 94.4 & QL, P1, Refold &\\
J2009+3326 & 1.438 & 263 & 71.1 & 0.1 & 0.82 & 3.0 & 0.15 & 32.4 & 23.9 & QL, Refold & Yes \\
J2010+3230 & 1.442 & 371 & 70.4 & --0.5 & 0.60 & 2.2 & 0.12 & 32.6 & 23.4 & QL, P1, Refold & Yes \\
J2011+3331 & 0.932 & 298 & 71.3 & --0.0 & 2.50 & 2.6 & 0.38 & 21.2 & 39.4 & QL, P1, Refold & Yes \\
J2018+3431 & 0.388 & 222 & 73.0 & --0.8 & 1.70 & 2.0 & 0.24 & 47.4 & 31.7 & QL, P1, Refold & Yes \\
\enddata
\tablecomments{A comprehensive list of all pulsars detected by the precursor survey as well as those we expected to 
detect, given their high $({\rm S/N})_{\rm th}$ quantities. We list each pulsar's period ($P$), dispersion measure 
(DM), Galactic longitude ($\ell$), Galactic latitude ($b$), angular offset from the closest beam ($\Delta\theta$) and
duty cycle ($\delta$), as well as $({\rm S/N})_{\rm th}$, $({\rm S/N})_{\rm meas}$. PALFA precursor data were run through
two processing pipelines, Quicklook and {\sc Presto 1} (described in \S\ref{sec:scNda}), then converted into 4-bit
files and stored. Pulsars detected by Quicklook (QL) or {\sc Presto 1} (P1) pipelines are marked accordingly; those detected
after refolding archived, 4-bit data files have ``Refold'' in the ``Pipeline Detected?'' column.
Previously unknown pulsars discovered by the precursor survey are marked with a ``Yes'' in the last column. For 
sources without an available flux density measurement, we did not compute $({\rm S/N})_{\rm th}$.
Previously determined parameters ($P$, DM, $\ell$, $b$ and flux density) were obtained from the ATNF Pulsar Catalog 
\citep{psrcat}. Missing parameters, $({\rm S/N})_{\rm th}$ and $({\rm S/N})_{\rm meas}$ for example, are denoted by 
dashes (---).}
\end{deluxetable*}

\subsection{Defining Detectability}\label{sec:defdet}
The PALFA multibeam receiver is composed of seven beams, each with an average full width half maximum (FWHM) of 
$\sim3.35\arcmin$; adjacent beams are separated by $\sim5.5\arcmin$, or $\sim1.6$ half-power beamwidths. Outer beams and the central beam have gains of 8.2 and \unit[10.4]{K Jy$^{-1}$} respectively \citep{cor06}. Although previous 
population studies have modeled gain patterns using Gaussian functions \cite[e.g.,][]{lor06}, we use an Airy disk 
function to better model the additional gain from the side lobes of individual beams. Although this is not a perfect 
representation of the PALFA survey's true gain pattern --- in fact, the side lobes of the outer ALFA beams are highly 
asymmetric (see Spitler et al., submitted, for a more precise model) --- the Airy disk captures Arecibo's off-axis 
gain better than the Gaussian model and still provides the simplicity required to run population simulations quickly. 

The theoretical signal-to-noise ratio $({\rm S/N})_{\rm th}$ for a given pulsar with flux density ($S_{1400}$) 
measured in mJy at \unit[1400]{MHz}, spin period $P$, and pulse width $W$ is given by
\begin{equation}
({\rm S/N})_{\rm th} =\frac{S_{1400}~G\sqrt{n_p~t_{\rm obs}~\Delta f}}{\beta~\Tsys}\sqrt{\frac{1-\delta}{\delta}},
\label{snr}
\end{equation}
where $\delta=W/P$ is the pulse duty cycle; $G$ is the gain in K~Jy$^{-1}$ of a specific beam, $n_p=2$ is the number 
of summed polarizations, $t_{\rm obs}$ is the integration time ($\unit[134]{s}$ and $\unit[67]{s}$ for inner- and 
outer-Galaxy observations, respectively), $\Delta f=\unit[100]{MHz}$ is the bandwidth, $\beta=1.16$ is a correction 
factor that accounts for losses in the digitization process and $\Tsys$ is the system temperature measured in K 
\citep{dew85}. Flux densities $S_{1400}$ were obtained from the ATNF Pulsar Catalog \citep{psrcat} for known pulsars
and \cite{nice13} for pulsars discovered by the PALFA precursor survey.
Equation \ref{snr} is an approximation since this treatment assumes top-hat pulse profiles and ignores the
considerable variability in pulse shape. The majority of pulsars have Gaussian-shaped profiles however, so this
approximation works well in most cases.

Hereafter $({\rm S/N})_{\rm th}$ will refer to theoretical signal-to-noise ratios, computed using 
Equation \ref{snr}, while $({\rm S/N})_{\rm meas}$ refers to signal-to-noise ratios measured from PALFA detections
as described in \S\ref{sec:snmeas} and specifically, Equation \ref{snrmeas}.

\begin{figure*}
\begin{center}
\includegraphics[scale=1.0,angle=0]{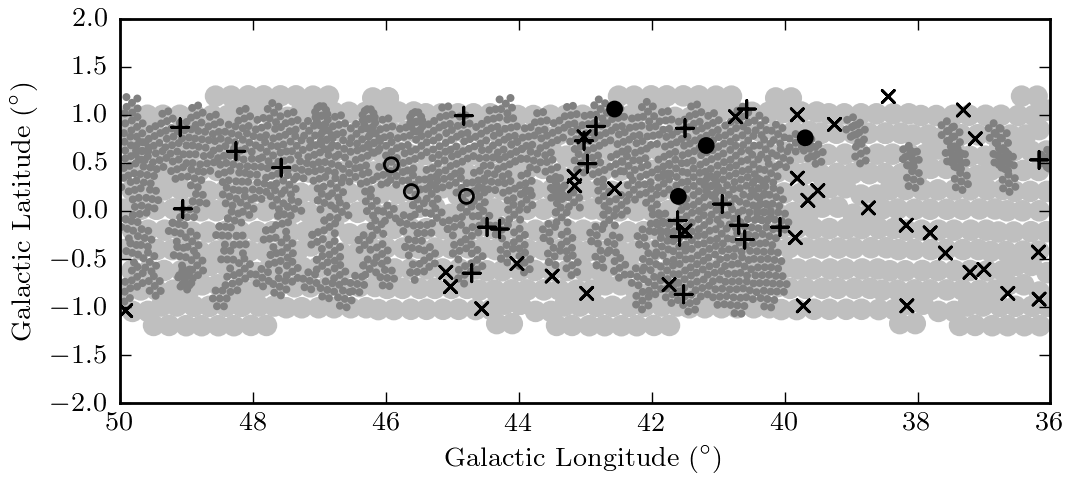}
\caption{Beam positions for the PMPS (light gray) and PALFA precursor survey (dark gray) are shown here with known 
pulsar positions superimposed. The Parkes beam radii are about 4 times as large as those of Arecibo; 
the points indicating beam positions have been scaled appropriately relative to one another. Only PMPS beams within 
1.2$^{\circ}$ of the Galactic plane are plotted since this more than covers the Galactic latitude limits of the 
PALFA precursor survey. Of the 58 previously known pulsars plotted here, many were too far from the nearest 
precursor survey beam center, making them undetectable (denoted by $\times${\it s}). Only 24 of 59 were deemed detectable, 
given the precursor survey's patchy coverage of this Galactic sector, and were considered in comparing the two surveys.
Known pulsars detected by the precursor survey are marked with $+${\it s}, while expected detections that were missed are marked with $\bigcirc${\it s}. Filled circles indicate the positions of PALFA precursor survey discoveries in the region overlapping with PMPS.}
\label{fig:overlap}
\end{center}
\end{figure*}

Since gain is a function of a source's angular offset from the beam center, we model it as an Airy disk so that the gain
\begin{equation}
G=G_0 \bigg(\frac{2~{\rm J_1}(k~a~\sin(\theta))}{k~a~\sin(\theta)}\bigg)^2,
\label{gain}
\end{equation}
where $J_1$ is a Bessel function of the first kind, $G_0$ is the maximum on-axis gain of the beam, 
$k = 2\pi/\lambda$ is the wavenumber  ($\lambda$, the observation wavelength), $a$ is the effective aperture radius 
($\sim\unit[220$]{m}), and $\theta$ is the angular offset of a source from the beam center, measured in radians. 
In predicting S/N for a given pulsar, the pulsed nature of its emission must be taken into account, as shown by the 
final term in Equation \ref{snr}. For all pulsars that were detected in the precursor survey, we computed $W_{\rm eq}$, then
$\delta$ as described in \S\ref{sec:snmeas}. For sources that were not detected, we divide the pulse width at 
half maximum (W50), from the ATNF Pulsar Catalog \citep{psrcat}, by the period to compute $\delta$, 
then $({\rm S/N})_{\rm th}$. Finally, $\Tsys$ includes the receiver 
temperature ($T_{\rm rec} = \unit[24]{K}$) and sky temperature ($T_{\rm sky}$), which varies as a function of 
position and frequency as shown by \cite{has82}. Since this sky temperature map describes $T_{\rm sky}$ at 
\unit[408]{MHz}, we convert these values into \unit[1.4]{GHz} sky temperatures using an assumed spectral index of 
$\alpha=2.6$, that is $T_{\rm sky} \propto \nu^{-\alpha}$.

Although there are many factors involved, we assume a 1:1 relationship between $({\rm S/N})_{\rm meas}$ and 
$({\rm S/N})_{\rm th}$ in order to use S/N as a prediction tool for 
the detectability of known sources. The true relationship between $({\rm S/N})_{\rm meas}$ and $({\rm S/N})_{\rm th}$ 
can be seen in Figure \ref{fig:snr}.

Using a complete list of beam positions, we found the survey observations carried out closest to known pulsars in 
the precursor region (i.e. minimizing angular offset, $\theta$). For each of these positions, we found the maximum 
expected gain for a given pulsar using Equation \ref{gain}. Previously measured parameters for known pulsars allowed 
us to compute a theoretical signal-to-noise, $({\rm S/N})_{\rm th}$, as shown in Equation \ref{snr}. We define a 
known pulsar to be detectable if we find $({\rm S/N})_{\rm th}>9$ for that pulsar. A full list of pulsars detected 
by the precursor survey as well as those considered detectable due to their $({\rm S/N})_{\rm th}$ values can be 
found in Table \ref{tab:full}. Before PALFA began, there were 84 known pulsars positioned inside the target precursor 
survey region, although this sky area was not covered uniformly; 31 of 84 were deemed detectable, while 33 were 
actually detected, and seven had no previous flux measurements. Of the 51 non-detections, most can be attributed 
simply to the sources not being close to a PALFA precursor survey beam pointing, as the survey had only limited 
coverage in this region. Figure \ref{fig:overlap} shows the portion of the precursor survey that overlaps with the 
PMPS), an example of this limited coverage. Three of the 51 non-detections (B1910+10, J1913+1145, and B1911+11) were 
unexpected, since $({\rm S/N})_{\rm th}>9$ for these sources; one of the 33 detections (B1929+20) was also unexpected, 
given its low $({\rm S/N})_{\rm th}$ value. The non-detections could be due to a variety of factors --- most likely 
RFI. Scintillation could have also suppressed the expected signal during precursor survey observations or boosted the 
signal during initial flux measurements. It is unlikely scintillation affected the detectability of J1913+1145, however, because of this source's high DM (\unit[637]{pc~cm$^{-3}$}). Given the short integration time near each of these sources (\unit[134]{s}), the pulse-to-pulse variability may have strongly affected $({\rm S/N})_{\rm meas}$ since 
relatively few pulses were recorded. Also, because of the large error bars on $({\rm S/N})_{\rm th}$ 
($\sim30\%$ fractional error) due to uncertainties in flux measurements, the sources may simply be weaker than expected.

Although most sources with high $({\rm S/N})_{\rm th}$ values were detected by the precursor survey's processing 
pipelines, five such sources were not.  For each of these cases, we employed the same procedure as introduced in Section
\ref{sec:snmeas}, using known periods and dispersion measures to dedisperse and fold the data from the closest pointing 
to each source.  For the three sources mentioned earlier (B1910+10, B1911+11, and J1913+1145), no pulsations were 
detected; for the other two, J1906+0649 and J1924+1631, pulsations are evident,but relatively weak. PSR J1906+0649 was 
likely missed because of the RFI environment at Arecibo.

In addition to the 33 re-detected pulsars in the region, PSR J1924+1631 was discovered shortly after the
precursor survey was completed, when the PALFA survey underwent an upgrade to a new backend with three 
times more bandwidth. This source was then retroactively found in precursor survey data with $({\rm S/N})_{\rm meas}$ 
just above the detection threshold and has therefore been included in analysis that follows. Strong RFI present in the 
refolded precursor data explains why this source was not discovered earlier. P. Lazarus (2014, in preparation) will 
describe the most recent processing pipeline in detail, 
address the RFI environment and its effect on the PALFA survey's ``true'' sensitivity.

\begin{deluxetable*}{lcc}[h!]
\tablecaption{\label{tab:popparams} Parameters Used in Population Simulations}
\tablecolumns{3}
\tablehead{Prior Distribution Parameter & Normal PSR Simulations & MSP Simulations}
\startdata
Luminosity & Log Normal: $\mu=-1.1$; $\sigma=0.9$ & Log Normal: $\mu=-1.1$; $\sigma=0.9$ \\
Period & Log Normal: $\mu=2.7$,$\sigma=-0.34$ & (see Figure \ref{fig:pdist}) \\
Radial & Gamma Function: \citep[see][]{lor06} & Gaussian: $\sigma=6.5$ kpc \\
Scale height & \unit[0.33]{kpc} & \unit[0.5]{kpc} \\
Duty Cycle & (explained in \S\ref{sec:genpdf}) & (explained in \S\ref{sec:genpdf}) \\
Electron Model & NE2001 & NE2001
\enddata
\tablecomments{Assumed parameter values/distributions for normal and millisecond pulsar populations respectively. 
These parameters are used as input values to the appropriate {\sc PsrPopPy} functions, which generate an underlying, 
synthetic population. Changing input parameters directly affects the number of detections expected from a given 
simulated survey.}
\end{deluxetable*}

\section{PMPS Overlap Region}\label{sec:overlap}

The PALFA precursor survey region overlaps the region covered by the PMPS in Galactic longitude, 
$36^{\circ}\lesssim \ell \lesssim 50^{\circ}$. Although there were 58 previously known pulsars in this longitude range
and within $\sim1^{\circ}$ of the Galactic plane when the precursor survey took place (see Figure \ref{fig:overlap}), 
we compare the PMPS and precursor
survey detections only based on sources deemed detectable by the precursor survey. We justify this criterion based 
on the fact that, due to patchy coverage, only $\sim$10\% of the overlap region
lies within an angular offset $\Delta\theta\sim1.2\arcmin$ of a precursor beam center. We choose $1.2\arcmin$ since this
is the average angular offset $\langle\Delta\theta\rangle={\rm FWHM}/2\sqrt{2}$ for the precursor survey. Half of
all sources that fall within a radius ${\rm R}={\rm FWHM}/2$ of the nearest beam center will also be within the average 
angular offset $\langle\Delta\theta\rangle$.

The PMPS discovered or detected all 24 of the previously known pulsars in this region considered detectable by the
PALFA precursor survey. The precursor survey detected 21 of these, and discovered an additional four sources in this
region. The PMPS retroactively detected two of these four precursor discoveries in archival data \citep[e.g.,][]{lor06}.

One of the three detectable known pulsars 
that the precursor survey missed, B1910+10, had a $({\rm S/N})_{\rm th}$ value of $\sim11$ 
(see Table \ref{tab:full}), just above the detectability threshold of $({\rm S/N})_{\rm th}=9$; the 
other two, J1913+1145 and B1911+11, were expected to be detected with $({\rm S/N})_{\rm th}=23$ and 36 respectively.
Error in $({\rm S/N})_{\rm th}$ is 
$\sim30\%$, which reflects the error in flux measurements and can easily explain the first non-detection. It 
is much harder to explain why J1913+1145 and B1911+11 were not detected, given their high $({\rm S/N})_{\rm th}$ values, but other systematics such as RFI and scintillation may explain these discrepancies.

\begin{figure}[b!]
\begin{center}
\includegraphics[scale=0.6,angle=0]{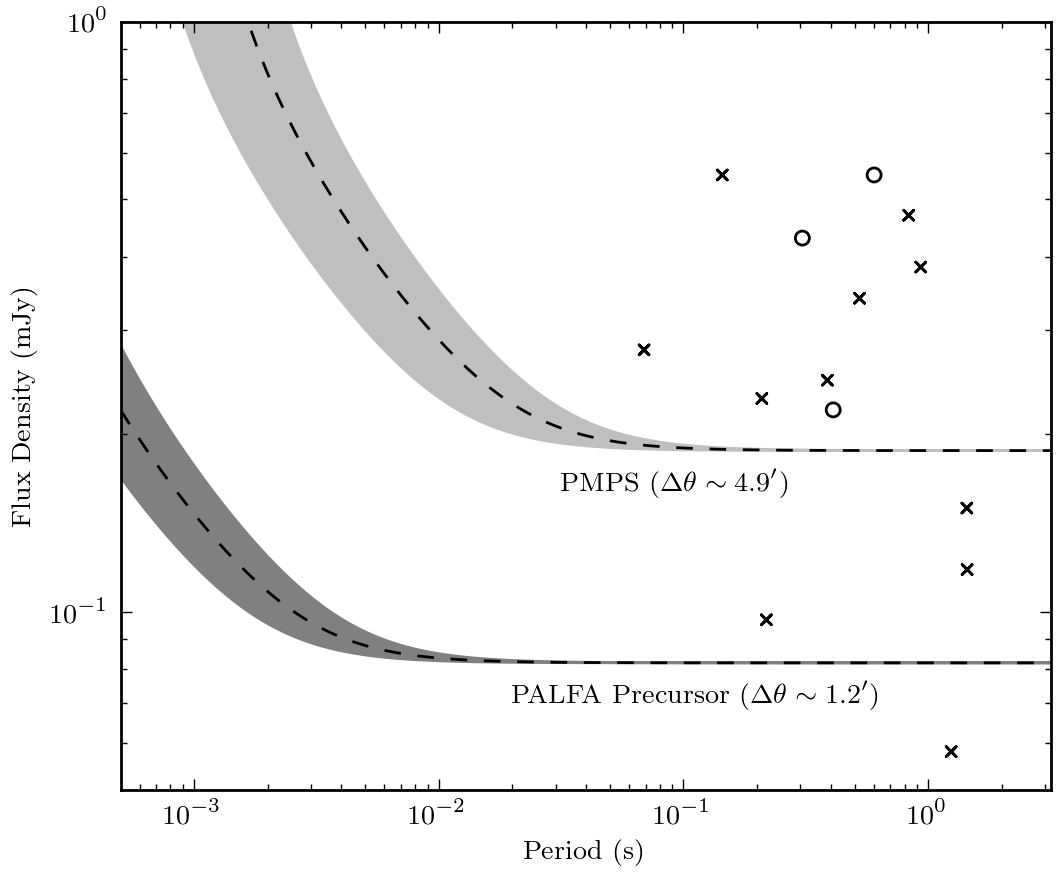}
\caption{Sensitivity as a function of period for the precursor survey is shown in dark gray; the PMPS curve 
(light gray) is shown here for comparison. The dashed lines in each case show the sensitivity to 
$\rm DM=\unit[100$]{pc~cm$^{-3}$} sources, while upper and lower limits of the shaded regions give minimum flux 
density sensitivity to pulsars with $\rm DM=150$ or \unit[$50$]{pc~cm$^{-3}$} respectively. These curves are plotted
using the average angular offset \avgdth~ between a source and a beam position. For a random distribution of pulsars
on the sky, $\sim50$\% should fall within an angle \avgdth~ from the nearest beam position. Precursor survey discoveries are
superimposed as $\times${\it s}, while expected detections B1910+10, B1911+11 and J1913+1145 that were missed by the
PALFA precursor survey, but detected by PMPS are shows with $\bigcirc${\it s}.}
\label{fig:sens}
\end{center}
\end{figure}

The four precursor survey discoveries --- J1901+0621, J1904+0738, J1905+0902 and J1906+0746 --- have relatively high 
dispersion measures and were all detected near the signal-to-noise threshold with $15<({\rm S/N})_{\rm meas}<22$, so 
it is not surprising that they were not detected by previous surveys. PSR J1906+0746  is a \unit[144]{ms} pulsar in 
a relativistic, \unit[3.98]{hr} orbit and was initially missed during manual inspection of PMPS candidate plots due 
to RFI with a period similar to that of the pulsar \citep{lor06b}. Both J1906+0746 and J1901+0621 were found 
retroactively in Parkes data, which was expected, given that both are moderately bright sources with flux densities 
at \unit[1400]{MHz} of about \unit[0.5]{mJy}. The other two discoveries, J1904+0738 and J1905+0902, are much 
fainter --- 0.23 and \unit[0.097]{mJy} respectively \citep{nice13}. These discoveries show preliminary evidence that 
with Arecibo's high sensitivity, the PALFA precursor survey probed a deeper and lower-luminosity pulsar population 
than previous surveys. However, the three unexpected non-detections suggest that the PALFA precursor survey did not
realize its full sensitivity and more work is required to better understand Arecibo's RFI environment and develop
mitigation techniques.

The relative sensitivity limits as a function of period and DM for the PMPS and precursor surveys are compared in 
Figure \ref{fig:sens}. To generate these curves, we used an average $T_{\rm sky}$ value for 
each survey region, assumed a constant pulse duty cycle of $\delta=0.05$, 
and applied the empirical pulse broadening function 
from \cite{bhat04} to account for multipath scattering in the interstellar medium. For the three objects that 
were detected at Parkes, but not in the PALFA precursor 
survey (B1910+10, J1913+1145 and B1911+11), all have periods between 300 and \unit[600]{ms}, a regime where the
PMPS nominal sensitivity limit in Figure \ref{fig:sens} is about twice as high as the precursor survey's. However, the angular
offsets to these sources ($6.6\arcmin$, $4.7\arcmin$ and $2.5\arcmin$ respectively for PMPS and
precursor values can be found in Table \ref{tab:full}) imply that both surveys were equally sensitive to them since
the PALFA precursor beam (FWHM$\sim3.35\arcmin$) is much narrower than that of the PMPS (FWHM$\sim14.4\arcmin$)
and its sensitivity therefore drops off more quickly as a function of $\Delta\theta$. Taking angular offsets into account,
B1910+10 ($S_{1400}=0.22$~mJy) falls below the adjusted minimum sensitivity limit (\unit[$\sim0.26$]{mJy} for both surveys), but
B1911+11 and J1913+1145 do not, so angular offsets alone do not explain
why these sources went undetected. Since other sources with lower flux densities and similar angular offsets {\it were}
detected (i.e. J0628+0909, J1906+0649, J1906+0912, J1907+0740, J1907+0918, J2011+3331), 
we conclude that transient effects such as RFI decreased the signal-to-noise ratios of B1910+10, 
B1911+11 and J1913+1145 and possibly scintillation for the former two.

\section{Population Analysis}\label{sec:pop}

The analysis presented here uses {\sc PsrPopPy} --- a package that models the Galactic population and evolution of 
pulsars. With this software, we populated a synthetic galaxy with pulsars whose attributes like cylindrical spatial 
coordinates, period, DM, luminosity, etc. were chosen from pre-determined PDFs \citep{lor06}. 
{\sc PsrPopPy}\footnote{https://github.com/samb8s/PsrPopPy} is a Python implementation 
of {\sc PSRPOP}\footnote{http://psrpop.sourceforge.net/}, which was written in Fortran \citep{lor06}; it shares 
much of the same functionality, but the object-oriented nature of Python and improved modularity of the code make 
it more readable and easier to write plug-ins for specific modeling purposes. Further details on the {\sc PsrPopPy} 
software package are forthcoming \citep{bates14}.

\subsection{Generating Pulsar Population PDFs}\label{sec:genpdf}

In order to deduce the sizes of the underlying Galactic normal and millisecond pulsar populations, we compared the 
results of {\sc PsrPopPy} simulations to the PALFA precursor survey's detection statistics for each of these two 
classes of pulsar. In each case, we made a set of assumptions about the underlying population (see Table 
\ref{tab:popparams}) and drew spatial and intrinsic pulsar parameters from assumed distributions to form a 
synthetic Galactic population. We simulated a survey of this synthetic population by computing 
$({\rm S/N})_{\rm th}$ as was discussed in \S\ref{sec:surv}. Again, detections were then defined as sources with 
$({\rm S/N})_{\rm th}>9$. The assumptions that went into our simulations, outlined in Table \ref{tab:popparams}, 
were largely drawn from the work by \cite{lor06} for the normal pulsar population. In that paper, however, the 
luminosity distribution for normal pulsars was assumed to behave as a power law with a low-luminosity cutoff of 
\unit[0.1]{mJy kpc$^2$}. Since the PALFA precursor survey's sensitivity dips below this cutoff value in some cases, 
we instead adopt a log-normal luminosity distribution, introduced by \cite{fgk06}. 

Since far fewer MSPs are known, we have very little information about the population's spatial and intrinsic 
parameter distributions, so some assumptions are simply adopted from the normal pulsar population (luminosity 
and radial distributions), while others are grounded in some preliminary experimental results (scale height, 
period and duty cycle distributions). In this case, we used a Gaussian radial distribution with a standard 
deviation of \unit[6.5]{kpc} and an exponential scale height larger than that of normal pulsars to reflect the 
fact that MSPs are distributed more uniformly across the sky. The Gaussian radial model for MSPs in the Galaxy 
is similar to that of normal pulsars, but makes no assumption about a deficiency of sources toward the Galactic 
center, an effect observed from full \emph{normal} pulsar population synthesis and modeled with a Gamma function 
\citep{lor06}.

\begin{figure}[b!]
\begin{center}
\includegraphics[scale=0.6,angle=0]{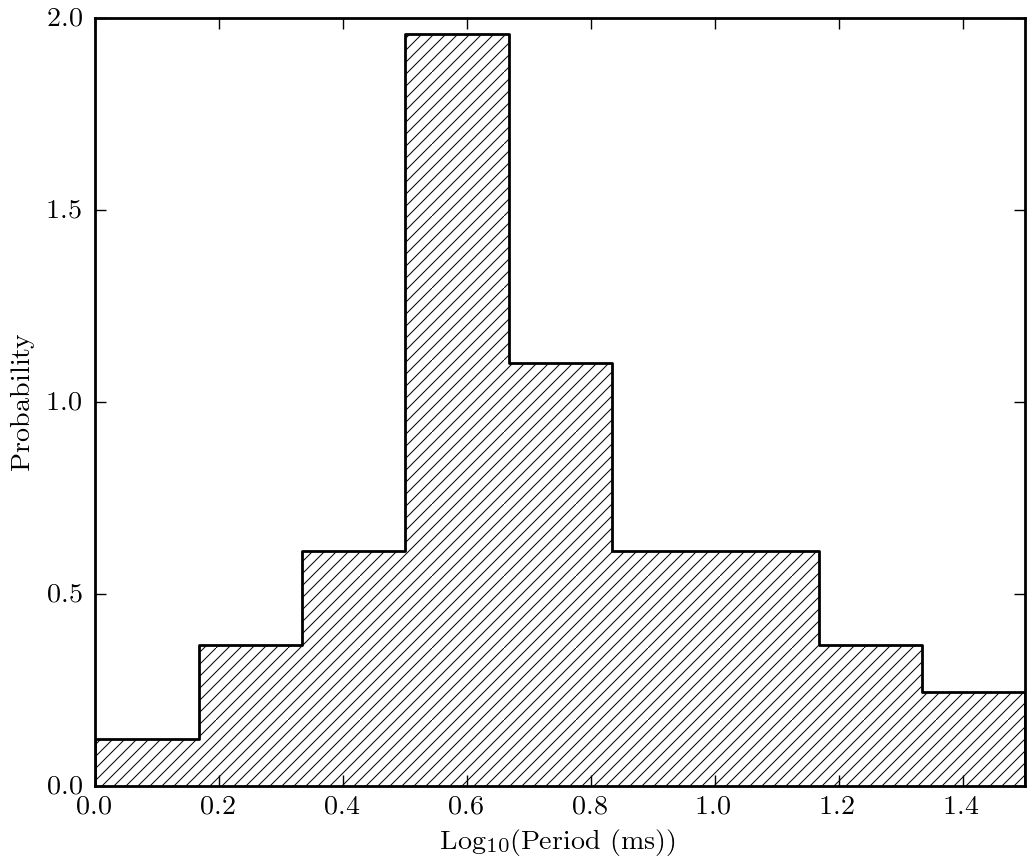}
\caption{This histogram shows the ad-hoc MSP period distribution used in simulations, which peaks at periods 
close to 3--ms. A more precise, empirically-based distribution is forthcoming and will be 
based on MSPs detected in the PMPS and HTRU surveys.}
\label{fig:pdist}
\end{center}
\end{figure}

We adopted the period distribution shown in Figure \ref{fig:pdist} from \cite{lor13}, where it was initially 
realized by adjusting the weights of various bins from a flat distribution (in $\log P$) until preliminary 
simulations matched the sample of observed MSPs from PMPS. Unlike normal pulsar duty cycles, which show inverse 
proportionality to the square root of spin period (i.e. shorter-period pulsars have wider pulses), MSPs tend to 
exhibit relatively constant duty cycle across period, with larger scatter about some mean value than the normal 
pulsar population \citep{kram98,smits09}. Therefore, our simulations assumed MSP duty cycles to be independent 
of period.

To make the simulated detections as realistic as possible, we used precursor survey parameters in 
signal-to-noise ratio calculations, and modified {\sc PsrPopPy} to accept the survey's true pointing positions, 
as well as corresponding integration times and specific beam gain values. For each population class, we 
performed simulated precursor surveys across a range of trial population sizes (85,000\---130,000 for normal 
pulsars and 5,000\---50,000 for MSPs). For each trial, we performed 2,000 simulated realizations of independent 
Galactic populations for MSPs and normal pulsars respectively. To form a likelihood function describing pulsar 
population size, we compared the results of these simulations to the true number of detections for each 
population class in the precursor survey. The precursor survey only detected a single MSP (B1937+21), so the 
likelihood was computed by dividing the number of simulations that resulted in a single detection by the total 
number of simulations at that population size.

\begin{figure*}
\begin{subfigure}{}
  \centering
  \includegraphics[width=0.49\linewidth]{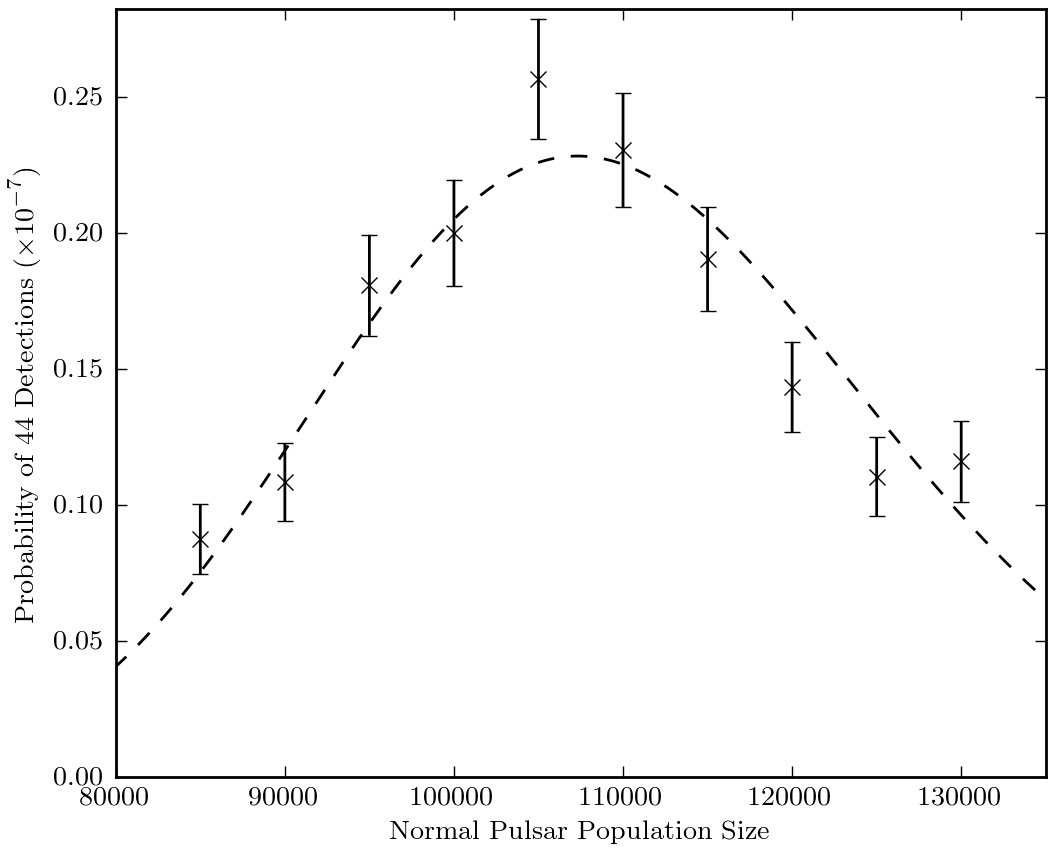}
\end{subfigure}
\begin{subfigure}{}
  \centering
  \includegraphics[width=0.5\linewidth]{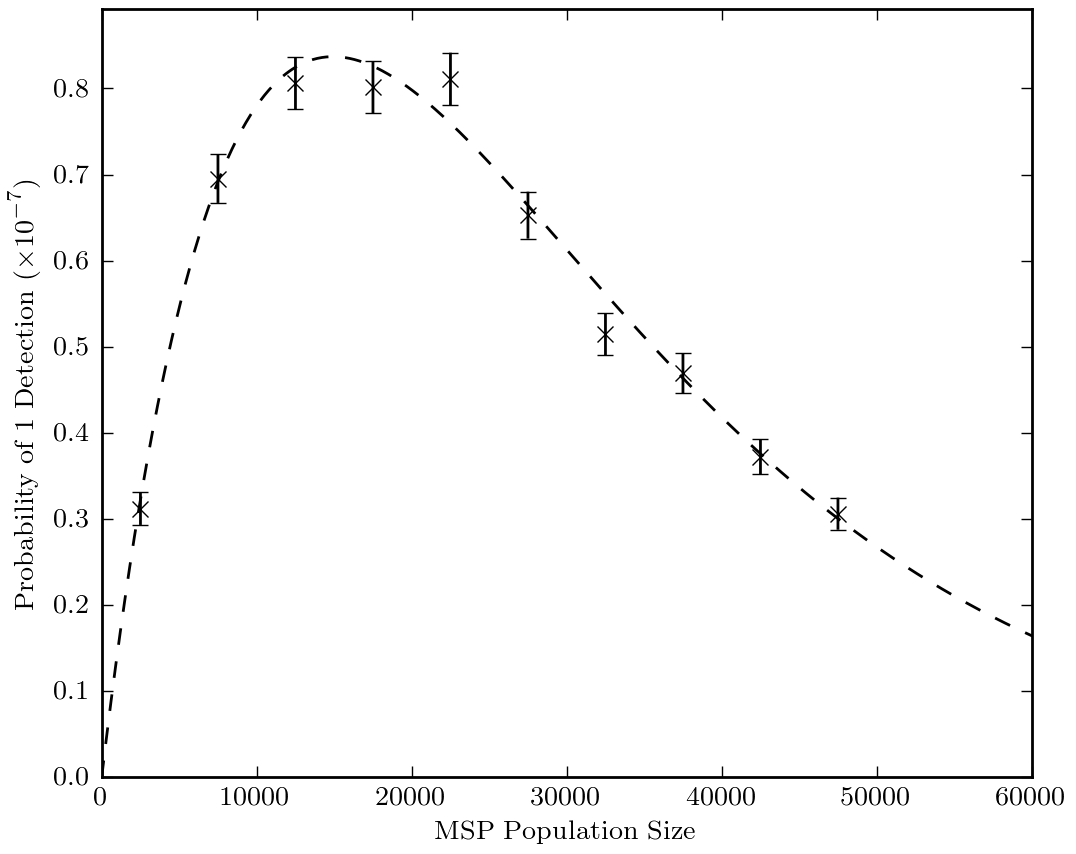}
\end{subfigure}
\caption{In each plot, black `\emph{x}'s show results of 2,000 population simulations at 10 different trial 
pulsar population sizes. The normal pulsar population PDF (left plot) was constructed with trial simulations 
using population sizes between 85,000 and 130,000 sources, while the MSP PDF (right plot) used between 5,000 
and 50,000 sources in trial simulations. In both cases the black dashed line shows a normalized binomial 
distribution fit to the data. Using these fits, we find that the most probable \emph{detectable} Galactic 
normal and millisecond pulsar population sizes are $\sim107,000$ and $\sim15,000$ respectively.}
\label{fig:pdfs}
\end{figure*}

Of the 45 detections listed in Table \ref{tab:full}, we exclude B1937+21 (MSP) from our normal pulsar analysis.
Although J1906+0746 is in a binary system, it is a young pulsar with a characteristic age of 112 kyr and has
likely not undergone recycling from its companion, so we include it in our analysis. The likelihood function was 
formed by dividing the number of simulations that detected 44 pulsars by the total number of simulations at a 
given trial population size. We fit binomial distributions to simulated likelihood functions for normal and MSP 
populations (shown in Figure \ref{fig:pdfs}) in order to smooth simulation results and provide integrable 
functions to determine confidence intervals. For an underlying population of size $N$, a given simulation has 
$n$ successes (detections) and $N-n$ failures (non-detections); these kinds of binary outcomes are nicely 
modeled by binomial distributions.

The binomial distributions provide the functional form
\begin{equation}
p(n|N,\theta)=\frac{N!}{n!(N-n)!}\theta^n(1-\theta)^{N-n},
\label{binomial}
\end{equation}
which describes the probability of drawing $n$ pulsars from a total population of $N$ given some detection 
probability $\theta$. To select the $\theta$ value that produces posterior PDFs that best match the simulated 
data, we chose the one that minimized $\chi^2$, computed by comparing simulated to expected population 
distributions. Finally, the posterior population size PDFs are normalized so that they could be used to quote 
confidence intervals. With some number $n_{\rm success}$ of successful realizations (simulations in which the 
target number of detections is reached), the Poissonian error is given by $\sqrt{n_{\rm success}}$. Data points 
that reflect the probability of detecting exactly the target number of pulsars at a trial population size and 
their error bars are multiplied by the same constant required to normalize the best fit PDF. After looking at 
multiple realizations of the simulated data presented in Figure \ref{fig:pdfs} and comparing the standard 
deviation of data points at each population size to assumed Poissonian error bar magnitudes, we determined 
that the Poisson model accurately reflects the uncertainties in population sizes.

By integrating the PDFs shown in Figure \ref{fig:pdfs}, we find the mode and 95\% confidence interval for the 
normal pulsar population size to be $107,000\substack{+36,000 \\ -25,000}$. We find a lower mode for the MSP 
population size, $15,000\substack{+85,000 \\ -6,000}$ and the high uncertainty in the corresponding 95\% 
confidence interval reflects the fact that our prediction depends on a single MSP detection in the precursor 
survey. These results describe the respective Galactic pulsar populations that are beaming towards Earth and 
errors on most likely population sizes account only for statistical uncertainties due to the limited number of 
detections in the PALFA precursor survey, not for other sources (e.g. uncertainties in scale height, luminosity 
distribution, electron density model, etc.).

The confidence interval that the precursor survey places on the normal pulsar population is consistent with 
earlier results; \cite{fgk06} predict $120,000\pm20,000$ detectable normal pulsars, also using a log-normal 
distribution to model the pulsar luminosity function. The predicted MSP population size is also consistent with 
previous estimates; the upper limit we find easily encompasses the population size prediction made by 
\cite{lev13}, although the lower limit quoted in that paper, $30,000\pm7,000$, is more constraining. Neither of 
these 95\% confidence intervals is tight enough to put strict constraints on normal or millisecond pulsar 
population sizes, but the consistency is encouraging and we expect the full PALFA survey to place much more 
stringent constraints on these populations when complete. 

\section{Results \& Discussion}\label{sec:rNd}

Using input parameters from Table \ref{tab:popparams} to generate a synthetic, Galactic normal pulsar population, 
we found that the PALFA precursor survey should be expected to detect $\sim40$ sources. Through periodicity 
searches, 43 were found, which indicates that current population parameters, initially determined using PMPS 
results, are already quite accurate and applicable to a variety of situations. As we mentioned in \S\ref{sec:defdet}, 
three sources that we expected to detect were not detected, but it is common for 
$({\rm S/N})_{\rm th}$ and $({\rm S/N})_{\rm meas}$ values to not match perfectly. Due to uncertainties in 
initial flux measurements, there can be as much as $\sim30\%$ fractional error in $({\rm S/N})_{\rm th}$. 
Referring again to Figure \ref{fig:snr}, we show a general trend towards a slope of unity when plotting 
theoretical versus measured S/N for the detections made by the precursor survey, but there is significant scatter 
in these comparisons. Scatter 
like this can be caused by scintillation, RFI, poor prior flux measurements or some combination of all of these.

The precursor survey discovered 11 pulsars, four of which fell inside the region overlapping PMPS, allowing us 
to directly compare their respective sensitivities. While PMPS detected almost three times as many sources in 
this region, this discrepancy was largely due to the differences in sky coverage --- PMPS covered this area 
uniformly, while the precursor survey had large blocks of coverage missing and slight gaps between pointings due 
to a ``sparse sampling'' technique. In fact, only $\sim25\%$ of the overlap region was covered by the precursor 
survey to a sensitivity greater than or equal to that of PMPS. Even so, the PALFA precursor survey discovered 
four pulsars that PMPS missed; two of these four were retroactively found by reanalyzing archival data but the 
others ( J1904+0738 and J1905+0902) have high dispersion measures and very low fluxes --- an encouraging, albeit 
small, piece of evidence that Arecibo's sensitivity gives PALFA a glimpse at fainter and more distant pulsars. 
Figure 3 in \cite{nice13} uses more recent PALFA discoveries to show further evidence of PALFA probing deeper 
than previous surveys as do recent discoveries mentioned in \cite{craw13}.

We simulated a range of Galactic pulsar populations --- both non-recycled and recycled --- of various sizes and 
used the PALFA precursor survey's detection statistics to place limits on normal and millisecond pulsar 
population sizes respectively. By comparing experimental results to simulations, we formed PDFs for normal and 
MSP population sizes, then integrated these PDFs to define confidence intervals.

Assuming the most probable normal and millisecond population sizes according to the simulations described in 
\S\ref{sec:genpdf} are correct, we ran 1,000 trials with the same distribution parameter assumptions for each 
population to determine the most likely number of detections by the beginning of 2014 and after PALFA is 
complete. Averaging the results of these 1,000 trials in each case, we determine a predicted number of detections, 
then quote errors that are directly proportional to the 95\% confidence limits from normal and millisecond pulsar 
population PDFs. Following this procedure, we expect the full PALFA survey to detect 
$1,000\substack{+330 \\ -230}$ normal pulsars (this includes previously known sources that are re-detected) and 
$30\substack{+200 \\ -20}$ MSPs. Identical estimation techniques predict that $490\substack{+160 \\ -115}$ normal 
pulsars and $12\substack{+70 \\ -5}$ MSPs should have been detected by the beginning of 2014, but at the time, 
PALFA had detected 283 normal pulsars and 31 MSPs, respectively\footnote{See 
http://www.naic.edu/$\sim$palfa/newpulsars for discoveries; re-detected sources are as yet unpublished.}.

The discrepancy between observed and predicted detection rates is notable for the normal pulsar population. Given 
the numbers quoted here, PALFA has currently detected just over 50\% of the expected number of normal pulsars, 
according to simulations. These simulations do not yet take into account the local RFI environment of the PALFA 
survey, which certainly plays a role in the perceived dearth of pulsar detections as of early 2014. Two pulsars 
that went undetected by both {\sc Quicklook} and {\sc Presto 1} pipelines in the precursor survey, J1906+0649 
and J1924+1631, provide evidence that initial processing techniques were not optimal and improvements are necessary.
In repeated simulations of precursor detections in the inner Galaxy region, we find $30-50\%$ of simulated, detectable sources
had S/N values between 9 and 15 (just above the detection threshold). In the precursor survey, only about 10\% of detections 
had $({\rm S/N})_{\rm meas}$ values in this regime. Although the precursor survey discovered mostly low flux density sources,
the fact that only a small fraction of detections were near the S/N threshold suggests that some sources were missed or assumptions that determine 
our sensitivity curves are not entirely correct.

A potential factor of two lower sensitivity to normal pulsars because of RFI would bring the 
survey yield and simulated population into agreement. The most recent PALFA survey pipeline will be described in depth 
by P. Lazarus (2014, in preparation) and that paper will also construct PALFA's ``true'' sensitivity curve, taking into account the RFI environment by 
injecting artificial signals of varying strength into real data. In future work, we will reprocess precursor 
survey data with the current pipeline to see if it improves the shortcomings of earlier versions (e.g. 
inconsistent detection statistics, noted in Table \ref{tab:full}).

The assumed radial distribution of pulsars in the Galaxy (see Table \ref{tab:popparams}) could also contribute to the discrepancy between 
expected (simulated) and true pulsar yields. Since the distribution is based on extrapolated results from the PMPS, which surveyed higher-populated 
regions of the sky, population density estimates for longitudes farther from Galactic center may be inaccurate. Over-estimated
pulsar population densities in the Galactic longitude range surveyed by the PALFA precursor survey could be a factor in the discrepancies
we find between expected and actual pulsar detections there. Future refinement of pulsar population models using PALFA results
will provide consistency checks for existing population model parameters.

We note that the current number of MSPs detected by PALFA is consistent with predictions, but this is not surprising, 
given the high uncertainties in our model due to the precursor survey only detecting one MSP. As the number of 
detections increases, future predictions will be far more constraining so that we can re-examine initial 
assumptions about the MSP population characteristics.

Future population studies with the complete PALFA survey will contribute substantially to current population models 
because of the Galactic longitude ranges covered and Arecibo's unrivaled sensitivity (especially in the millisecond 
pulse period regime). As the number of normal and millisecond pulsar detections increases, our ability to refine 
specific, simulated model parameters that describe each underlying population will improve significantly.

\section{Acknowledgements}\label{sec:ack}
The Arecibo Observatory is operated by SRI International under a cooperative agreement with the National Science Foundation
(AST\---1100968), and in alliance with Ana G. M\'{e}ndez-Universidad Metropolitana, and the Universities Space Research
Association. MAM and JKS are supported through NSF PIRE award \#0968296. DJN is supported through NSF grant \#0647820.  VMK was supported by an NSERC Discovery and Accelerator Grant, the
Canadian Institute for Advanced Research, a Canada Research
Chair, Fonds de Recherche Nature et Technologies, and the
Lorne Trottier Chair in Astrophysics. JWTH acknowledges funding from NWO and ERC. Work at Cornell was supported by NSF Grants 
\#0507747 and \#1104617 and made use of 
the Cornell Center for Advanced Computing. Pulsar research at UBC is supported by an NSERC Discovery Grant and 
Discovery Accelerator Supplement and by the Canada Foundation for Innovation. PL acknowledges support 
of IMPRS Bonn/Cologne and NSERC PGS--D.



\end{document}